# Deciphering the Enigma of Wave-Particle Duality

Mani Bhaumik[1]

Department of Physics and Astronomy,
University of California, Los Angeles, USA 90095

A reasonable explanation of the confounding wave-particle duality of matter is presented in terms of the reality of the wave nature of a particle. In this view a quantum particle is an objectively real wave packet consisting of irregular disturbances of underlying quantum fields. It travels holistically as a unit and thereby acts as a particle. Only the totality of the entire wave packet at any instance embodies all the conserved quantities, for example the energy-momentum, rest mass, and charge of the particle, and as such must be acquired all at once during detection. On this basis, many of the bizarre behaviors observed in the quantum domain, such as wave function collapse, the limitation of prediction to only a probability rather than a certainty, the apparent simultaneous existence of a particle in more than one place, and the inherent uncertainty can be adequately understood. The reality of comprehending the wave function as an amplitude distribution of irregular disturbances imposes the necessity of acquiring the wave function in its entirety for detection. This is evinced by the observed certainty of wave function collapse that supports the paradigm of reality of the wave function portrayed in this article.

1.  Introduction

Ever since the advent of the wildly successful quantum theory nearly a century ago, its antecedent, wherein the wave aspect is always associated with a particle has been a source of great mystery to professional scientists and to the public as well. Louis de Broglie, who offered the hypothesis, considered the wave to be a fictitious pilot wave that guides a particle. Almost a quarter of a century later, David Bohm came up with the notion of a quantum potential instead, providing the quantum wave that guides the classical particle. He never expounded, however, on the source of energy of the quantum potential. Nonetheless, due to other appealing aspects of Bohm's theory such as the introduction of the concept of nonlocality, a small segment of the scientific community still seem to grant it credence.

[1] e-mail: bhaumik@physics.ucla.edu

Niels Bohr, the principal architect of the Copenhagen interpretation, was content to accept the duality paradox as an elementary aspect of the natural world. In his view, a quantum object will exhibit its wave aspect at some times and its particle nature at others, depending on the circumstances. He considered such duality an essential feature of complementarity, which he presumed to be an inherent property of nature.

During all this time, buoyed by the phenomenal success and the superb predictive power of the quantum theory, most practicing physicists avoided the conundrum by treating the wave function as merely a fictitious mathematical construct to be used for the algorithm called quantum theory.

This caused the pioneering proponent of the quantum theory, Albert Einstein, great consternation. He posited, "This double nature of radiation (and of material corpuscles)... has been interpreted by quantum-mechanics in an ingenious and amazingly successful fashion. This interpretation...appears to me as only a temporary way out..." [1]

"At the heart of the problem," Einstein said of quantum mechanics, "is not so much the question of causality but the question of realism." [2, p.460] Einstein's insistence on the reality of the wave function, however, remained conspicuously dormant until recently, when it became the subject of lively discussion in the form of $\psi$-ontic versus $\psi$-epistemic debate. A comprehensive review has been provided by M. Leifer [3]. It would appear that $\psi$-ontic theories advanced by Colbert and Renner [4] and Pusey, Barrett, and Rudolf [5], which advocate the objective reality of the wave function, are gaining more traction. Their presentation, however, is based primarily on an information-theoretic viewpoint and does not take into account physical aspects such as the conservation of energy or momentum.

In this communication, we provide a paradigm of the reality of the wave function based on energy-momentum considerations that cogently explicates the enigma of the wave-particle duality, an enigma which is inseparably connected with several other conundrums of quantum physics, such as, Heisenberg's uncertainty principle, the Born rule of probability, collapse of the wave function, and the apparent simultaneous existence of a particle in more than one place. The concept of the objective reality presented here solely in terms of physical parameters is more explicit, albeit for a single particle, and ought to harmonize with the information-theoretic approach.

The topic of wave-particle duality itself has a fascinating history. Notably, it was Einstein himself who fostered this innovative notion by advocating the real existence of quanta of radiation or photons. Previously Maxwell and others had quite convincingly established the wave nature of electromagnetic radiation. An abundance of experiments on the interference, diffraction, and scattering of light had substantiated it beyond any reasonable doubt. Thus it was greeted with utter shock and disbelief when Einstein [6] argued in 1905 that under certain circumstances light behaves not as continuous waves but as discontinuous, individual particles. These particles, or "light quanta," each carried a "quantum," or fixed amount, of energy.

In the face of the almost unanimous opposition of his peers, Einstein remained perhaps the principal champion of the wave-particle duality of radiation for almost two decades, until he was finally vindicated in 1923 by the spectacular observation of the particle aspect of x-rays in the Compton Effect. The following year, de Broglie extended the idea of wave-particle duality to matter particles with enthusiastic support from Einstein.

Soon the evidence for the matter wave came along with the apparent accidental discovery of electron waves by Clinton Davisson and Lester Germer in observing a diffraction pattern in the beam of electrons scattered by nickel crystal. [7] Shortly before that, G.P. Thompson and A. Reid also provided some evidence of matter waves by detecting a diffraction pattern when electrons passed through a very thin metallic foil [8]. Both Davisson and Thompson shared the Nobel Prize for their startling discovery, ushering the age of quantum physics in earnest.

In the meantime Schrödinger, inspired by de Broglie and Einstein" [9], formulated the wave mechanics of quantum physics, replacing the particle in classical mechanics with a wave function.

## 2. Reality of the Wave Function

In previous articles, [10], we presented a credible argument in favor of the existence of an objective reality underlying the wave function at the core of quantum physics. A synopsis is presented here. The ontology of the wave function advocated in this paper is primarily grounded on the incontrovertible physical evidence that all electrons in *the universe* are indistinguishable.

The answer to the long standing puzzle of why all electrons in every respect are universally identical, a feature eventually found to be true as well for all the other fundamental particles, was finally provided by the Quantum Field Theory (QFT) of the Standard Model of particle physics constructed by combining Einstein's special theory of relativity with quantum physics, a science which has evolved from his own pioneering contributions.

Nobel Laureate in Physics, Steven Weinberg [11], declares that QFT is an unavoidable consequence of the reconciliation of quantum mechanics with special relativity. It has successfully explained almost all experimental observations in particle physics and correctly predicted a wide assortment of phenomena with impeccable precision. By way of many experiments over the years, the QFT of Standard Model has become recognized as a well-established theory of physics. Another Nobel Laureate David Gross asserts [12] that all the pieces of the puzzle of the Standard Model of particle physics fit beautifully in QFT of the Standard Model such that there are no more pieces of the puzzle left to fit. Yet another Physics Nobel Laureate Franck Wilczek underscores [13, p. 96], "…the standard model is very successful in describing reality—the reality we find ourselves inhabiting." Expressions of such confidence encourage us to anchor our reliance on the QFT of the standard model.

One might argue that although the Standard Model accurately describes the phenomena within its domain, it is still incomplete as it does not include gravity, dark matter, dark energy, and other phenomena. However, because of its astonishing success so far, whatever deeper physics may be necessary for its completion would very likely extend its scope without repudiating its current depiction of fundamental reality.

According to QFT, the fundamental particles which underpin our daily physical reality are only secondary. Each fundamental particle, whether it is a boson or a fermion, originates from its corresponding underlying quantum field [14--17]. The particles are excitations of quantum fields possessing propagating states of discrete energies, and it is these fields which constitute the primary reality. For example, a photon is a quantum of excitation of the photon field (aka electromagnetic field), an electron is a quantum of the electron quantum field, and a quark is a quantum of the quark quantum field, and so on for all the fundamental particles of the universe.

By far, the most phenomenal step forward made by quantum field theory lies in the stunning prediction that the primary ingredient of *everything* in this universe is present in *each element of spacetime* (*x*, y, z, *t*) of this immensely vast universe [13, p.74]. These ingredients are the underlying quantum fields. We also realize that the quantum fields are alive with quantum activity. These activities have the unique property of being completely spontaneous and utterly unpredictable as to exactly when a particular event will occur [13, p.74]. Furthermore, some of the quantum fluctuations occur at *mind-boggling speeds* with a typical time period of $10^{-21}$ seconds or less. In spite of these wild infinitely dynamic, fluctuations, the quantum fields have remained immutable, as evinced by their Lorentz invariance, essentially since the beginning and throughout the entire visible universe.

3. **Quantum Particle in Motion**

As elucidated above, an electron represents a propagating discrete quantum of the underlying electron field. In other words, an electron is a quantized wave (or a ripple) of the electron quantum field, acting as a particle because of its well-defined energy, momentum, and rest mass, which are conserved fundamentals of the electron. However, even a single electron, in its reference frame, is never alone. It is unavoidably subjected to the perpetual fluctuations of all the quantum fields.

When an electron is created instantaneously from the electron quantum field, its position would be indefinite since a regular ripple with a very well defined energy and momentum is represented by a non-localized periodic function. However, the moment the electron comes into existence, quantum fluctuations facilitates its interaction with all the other quantum fields. For example, the presence of the electron creates a disturbance in the electromagnetic or the photon quantum field. Assisted by a fleeting quantum fluctuation, the disturbance in the photon field can momentarily appear as what is commonly known as a spontaneously emitted virtual photon. It is these interactions that allow the particle's position to be localized, which will now be described in more detail.

We understand how the quantum fluctuations continually and prodigiously create virtual electron-positron pairs in a volume surrounding the electron. "Each pair passes away soon after it comes into being, but new pairs are consistently boiling up to establish an equilibrium distribution.''[18] Even though each pair has only a fleeting existence, on an *average* there are a significant enough number of these pairs to generate a remarkably sizable screening of the bare charge of the electron.

In the same way, although any individual disturbances in the fields or the virtual particles due to quantum fluctuations have an ephemeral existence, there ought to be an equilibrium distribution of such disturbances present at any particular time affecting other aspects of the electron as well. The effect of these disturbances is very well established in phenomena such as the Lamb shift and the anomalous gyromagnetic factor of the electron's spin.

The electron's spin g-factor has been measured to a precision of better than one part in a trillion, compared to the theoretically calculated value that includes QED diagrams up to four loops [19] Therefore it would be reasonable to assume that the average number of disturbances of all quantum fields present at any particular time will be strikingly stable in spite of their flitting in and out of existence.

Recall again that an electron is a quantized ripple of the electron quantum field, acting as a particle because it travels with its conserved quantities always maintained holistically as a unit. However, *due to interactions of the particle with all the other quantum fields, substantially equivalent to those involved in the Lamb shift and the observed spin g-factor*, the ripple in fact becomes very highly distorted immediately after its creation since the quantum fluctuations causing the interactions of the quantum fields have a typical time period of $10^{-21}$ second.

As soon as this happens, the electron itself ceases to be a normal particle and becomes more like a general disturbance consisting of irregular disturbances of all the quantum fields to various degrees. *However, it is imperative to comprehend that according to the edicts of QFT, only the combination of all the irregular disturbances at all times add up to the precise amount of energy, momentum, and mass of the electron.* In this process, it should be obvious that the electron ceases to be a ripple of single frequency and becomes a highly deformed *localized* travelling pulse.

It is well known that such a pulse, no matter how deformed, can be expressed by a Fourier integral with weighted linear combinations of simple periodic wave forms like trigonometric functions, mentioned by the author in earlier communications [20]. *The result would be a wave packet or a wave function that represents a fundamental objective reality of the universe.* Such a wave function would be smooth and continuously differentiable, especially using imaginary numbers if necessary in the weighted amplitude coefficients. The wave function ψ(x) will be given by the Fourier integral,

$$\psi(x) = \frac{1}{\sqrt{2\pi}} \int_{-\infty}^{+\infty} \emptyset(k)\, e^{ikx} dk \qquad (1)$$

where $\emptyset(k)$ is a continuous function that determines the amount of each wave number component k = 2π/λ that gets added to the combination.

From Fourier analysis, we also know that the spatial wave function ψ(x) and the wave number function $\emptyset(k)$ are a Fourier transform pair. Therefore we can find the wave number function through the Fourier transform of ψ(x):

$$\emptyset(k) = \frac{1}{\sqrt{2\pi}} \int_{-\infty}^{+\infty} \psi(x)\, e^{-ikx} dx. \qquad (2)$$

Thus the Fourier transform relationship between ψ(x) and $\emptyset(k)$, where x and k are known as conjugate variables, can help us determine the frequency or the wave number content of any spatial wave packet function.

### 4. Time Evolution of ψ(x)

In order to determine the time evolution of the wave packet function, we need to incorporate the time term to the spatial function. Accordingly,

$$\Psi(x, t) = \frac{1}{\sqrt{2\pi}} \int_{-\infty}^{+\infty} dk\, \emptyset(k)\, e^{i(kx - \omega(k)t)} \qquad (3)$$

We introduced ω (k) since the angular frequency will be quite often a function of the wave number k. The wave packet function has a dominant central wave number $k_0$ and a range of additional wave numbers on either side that combines to provide the necessary localization of the packet

The kinematics of the wave packet will conform always to the relativistic energy relation,

$$E^2 = m_0^2\ c^4 + p^2\ c^2 \qquad (4)$$

Or equivalently in terms of the Plank – Einstein formula

$$E = h\upsilon = mc^2, \qquad (5)$$

$$h^2\nu^2 = h^2\nu_0^2 + (pc)^2. \qquad (6)$$

Before the Electro-Weak symmetry breaking about a trillionth of a second after the big bang and the attendant manifestation of the Higgs field, all the wave packets representing the various particles having no mass, but a momentum $p = \hbar$ k, were speeding along with the velocity of light c, since the group velocity of the wave-packet

$$\upsilon_g = \frac{d\omega}{dk} = \frac{d(kc)}{dk} = c.$$

A sweeping change occurred in the kinematics of the wave packets after their interaction with the Higgs field, when a wave packet is considered to have acquired a mass, more specifically the rest mass $m_0$, [21] thereby reducing its translational motion. In other words, by interacting with the Higgs field, the wave packet, paraphrasing Einstein [22], has acquired inertia proportional to its energy content. By this process, different wave packets representing different particles acquire their rest masses, which is a measure of the strength of their coupling with the Higgs field.

Since the temperature and energy of the universe following the manifestation of the Higgs field were still very high compared to the rest mass energy, the kinematics of the wave packet (particle) obeyed the relativistic energy-momentum equation (4) or equivalently (6). However, the mean free path of any particle was rather small because of the rapid rate of pair production and annihilation inside the predominantly high energy photon gas.

Eventually, when the universe was about a few seconds old, it cooled down sufficiently below the threshold of all pair productions and nearly one in ten billion particles survived over the antiparticles as a result of asymmetry in bariogenesis during the early universe. The electrons were the last to escape pair production and annihilation in the primordial soup. Ultimately, the wave packets (matter particles) combined to form atoms and other forms of matter.

*Our goal is to show that the form of a wave packet given in eqn. (3) representing a particle holds well even when particles acquire mass. In this way, a particle persists as a wave packet from very shortly after its inception from the quantum field to its utilization in the formation of matter or in some sort of detection when the wave function collapses.*

## 5. Modulation of the Wave Function

Without any kinetic energy, a wave packet with the intrinsic energy $h\nu_0$ would become a standing wave packet corresponding to rest mass energy $m_0 c^2$. *The essential feature here is to recognize that even though the particle is at rest, its rest mass energy is not. It ought to manifest in the vibrations of some sort of a standing wave obeying the equ*ation,

$$\Psi(x, t) = W(x) e^{-iE_0 t/\hbar} \qquad (7)$$

The quantum of energy $h\nu_0$ corresponding to the rest mass is vibrating with a very significantly high frequency. Because of the substantially high frequency of the standing wave, when it starts to move even with a small velocity, special relativistic effects become manifest in its reference frame $S'$ as observed from the laboratory frame S. This is a concept brought forward from de Broglie's original thoughts by some recent authors in their treatment of the de Broglie wave. A cogent presentation has been advanced by Shanahan [23]. However, he had to propose a model particle with a standing wave packet at rest. In this presentation the wave packet is revealed to be a natural feature from its very origin.

With a boost velocity $v$ in the $x$ direction and applying the Lorentz transformations

$$x' = \gamma (x - v t),$$

$$t' = \gamma (t - \frac{vx}{c^2}),$$

where $\gamma$ is the Lorentz factor

$$\gamma = \frac{1}{\sqrt{1 - v^2/c^2}}$$

equation (7) for the standing wave packet becomes

$$\Psi(x, t) = W\left(\gamma(x - vt)\right) e^{-i\, m_0 c^2\, \gamma\left(-\frac{vx}{c^2}+t\right)/\hbar}$$

$$= W\left(\gamma(x - vt)\right) e^{(i\gamma m_0\, vx/\hbar)-(i\gamma\, m_0\, c^2 t/\hbar)}$$

$$= W\left(\gamma(x - vt)\right) e^{(ipx/\hbar)-(iEt/\hbar)} \quad (8)$$

Comparing equation (8) with that of a transverse wave, we can easily identify the wave number $k = p/\hbar$ as the postulated de Broglie wave number with the wave length $\lambda = h/p$. Equation (8) shows that the standing wave packet is now a Lorentz shifted wave packet moving with velocity $v$ and whose space phase is modulated by the complex quantity $e^{ipx/\hbar}$ that involves the momentum $p = mv$.

From the above investigations, it is clear that the well-known de Broglie wave length $\lambda$ associated with a particle is not really an independent wave but is seen as such due to relativistic effects producing a phase modulation of the wave packet that caries the energy.

## 6. Group Velocity of the Wave Packet

The following analyses ascertain that the group velocity of the wave packet is the translational velocity $v$. Because of the involvement of the velocity $v$, it is more convenient for this purpose to use the Einstein's energy, momentum relations rather than the equivalent Planck formulae.

The group velocity $v_g = \dfrac{\partial \omega}{\partial k} = \dfrac{\partial E}{\partial p} = \dfrac{\partial E}{\partial v}\left(\dfrac{\partial p}{\partial v}\right)^{-1}$ since $p = \hbar k$.

$$\frac{\partial E}{\partial p} = \frac{\partial}{\partial v}\left(\frac{m_0 c^2}{\sqrt{1 - v^2/c^2}}\right) \left[\frac{\partial}{\partial v}\left(\frac{m_0 v}{\sqrt{1 - v^2/c^2}}\right)\right]^{-1}$$

Using the quotient and chain rules of differentiation,

$$v_g = \frac{m_0\, v}{(1 - v^2/c^2)^{3/2}} \left[\frac{m_0}{(1 - v^2/c^2)^{3/2}}\right]^{-1}$$

$$= v$$

υ being the velocity of a "particle." From equation (8), the wave-packet moving with the group velocity υ and representing the particle of same velocity is described by the familiar wave function:

$$\Psi(x, t) = \psi(x, t)\, e^{i(kx - \omega t)} \tag{9}$$

where $\psi(x, t) = W[\gamma(x - \upsilon t)]$, $k = p/\hbar$, and $\omega = E/\hbar = 2\pi\gamma\nu_0$

and k varies with velocity for a massive particle.

Although the analysis presented above is conducive for an intuitive understanding of the phenomenon, the description of a particle with v=0, seems problematic since a quantum particle is never at rest due to its characteristic quantum jitter having a zero-point energy. The nature of the standing wave and how it is tangibly sustained also need further scrutiny. Therefore we present an alternative approach that circumvents these concerns.

When a particle acquires mass after the manifestation of the Higgs Field, the group velocity of the wave packet in eqn. (3) is no longer equal to the constant c, but a variable velocity v depending upon its energy-momentum. Consequently the spacetime dependence $e^{i(kx - \omega(k)t)}$ must change frequently, which can be elegantly deduced using the four-vector procedure of the special theory of relativity.

The transformation of the momentum four-vector ( $E/c$, $\vec{p}$ ) and the wave four-vector ( $\omega/c$, $\vec{k}$ ) that keeps their magnitude invariant is the Lorentz transformation. Considering a laboratory frame $S$ and a rest frame $S'$ with a boost velocity $\upsilon$ in the x direction, the Lorentz transformation relations for the momentum four-vector and the wave four-vector are:

$$E'/c = \gamma(E/c - \beta p_x) \quad \text{and} \quad \omega'/c = \gamma(\omega/c - \beta k_x)$$

$$p'_x = \gamma(p_x - \beta E/c) \qquad\qquad k'_x = \gamma(k_x - \beta\omega/c)$$

$$p'_y = p_y \qquad\qquad\qquad\qquad k'_y = k_y$$

$$p'_z = p_z \qquad\qquad\qquad\qquad k'_z = k_z \tag{10}$$

Where $\beta = \upsilon/c$ and $\gamma = 1/\sqrt{1 - \beta^2}$

To determine the proportionality constant between the two four-vectors, we compare their timelike components after multiplying the component of the wave four-vector by $\hbar$:

$$E'/c = \gamma(E/c - \beta\, p_x) \qquad (11)$$

$$\hbar\,\omega'/c = \gamma(\hbar\omega/c - \beta\,\hbar k_x) \qquad (12)$$

Subtracting eqn. (12) from eqn. (11), we have

$$E'/c - \hbar\,\omega'/c = \gamma(E/c - \hbar\,\omega/c - \beta\,p_x + \beta\,\hbar k_x) \qquad (13)$$

*Since* the laws of physics are the same in all inertial frames of reference, the Planck's law $E = \hbar\omega$ in frame $S$ should hold true in frame $S'$, giving us $E' = \hbar\omega'$. Thus eqn. (13) reduces to

$$\gamma\beta(p_x - \hbar k_x) = 0 \qquad (14)$$

*According to the zero product property of algebra, either $\gamma\beta = 0$ or $p_x - \hbar k_x = 0$.*

*Because of the quantum jitter, we can clearly say that $\gamma\beta$ is non zero. Then we have*

$$p_x = \hbar k_x \qquad (15)$$

Or more generally, $p = \hbar k$ irrespective of the mass of the particle, zero or otherwise. This relationship is the celebrated de Broglie hypothesis. However, we now realize that the relationship can indeed be derived and does not need to be a hypothesis.

As presented earlier, using the relationship in eqn.(15) the group velocity of the wave packet $v_g$ equals $v$, which is the velocity of the "particle" represented by the wave packet. Therefore, the wave packet moves with the velocity of a massive as well as a massless particle. Steven Weinberg also confirms this result using a slightly different consideration [24].

### 7. Kinematics of a Quantum Particle

Since a particle like an electron in motion is represented by a wave function as given by the equation (3), its kinematics cannot be described by the classical equations of motion. Instead, it requires the use of an equation like the

Schrödinger equation, which for a non-relativistic particle is given by

$$i\hbar \frac{\partial \psi(x,t)}{\partial t} = -\frac{\hbar^2}{2m} \nabla^2 \psi(x,t) + V\psi(x,t) \quad (11)$$

where $V$ is the classical potential.

We must now explore what exactly the wave function $\psi(x,t)$ actually represents. It would be tempting to think, as in fact Schrödinger originally did, that the wave function represents a smeared out particle. But recalling how the wave packet came in to existence it would be obvious that this is not the case. The wave packet consists of irregular disturbances and only the sum total of which represents the mass, energy-momentum, charge of a particle like electron. Therefore, the wave function is in fact a function of probability amplitudes for finding the particle.

It should be noted, however, that although the attributes of the various irregular disturbances are mostly characteristics of their respective quantum fields with different charge, spin, etc., they have one aspect in common. The element of disturbance in energy is identical for all fields. Energy density of a wave is given by the square of its amplitude. Therefore to get the probability density, we have to take the square of the amplitude of the wave function, which usually involves a complex quantity. Thus the square of the amplitude $\psi^*(x,t)\,\psi(x,t)$, should represent the probability density $P(x,t)$ for finding a particle in position space at time t. This is acknowledged as the renowned Born's rule, which is a necessary hypothesis of quantum mechanics. But as we have just presented, it is a natural consequence of the reality of the wave function revealed in this article.

However, it is of critical importance that the wave function is normalized:

$$\int_{-\infty}^{+\infty} \Psi^*(x,t)\,\Psi(x,t)dx = 1$$

Since the sum of all the probabilities has to be 1 for a single particle.

The wave function evolves impeccably in a unitary manner. But when the particle inevitably interacts with a classical device like a measuring apparatus, the wave function undergoes a sudden discontinuous change known as the wave function collapse. Although it is an essential postulate of the Copenhagen interpretation of quantum mechanics, the phenomenon has long been *perplexing* to the physicists [25]. However, a behavior like this would be a natural consequence of the distinctive nature of a quantum particle described in this article. In a measurement, since the holistic wave packet only in its totality always contains the conserved quantities of a particle like an electron such as its energy-momentum, charge, spin etc., it must be taken all at once or not at all. In other words, for measurement, the collapse of the wave function is essential since the entire wave packet holistically representing the particle has to be commandeered.

Parts of the wave function that might spread to a considerably large distance can also terminate instantaneously by the process involved in a credible quantum mechanical Einstein-Rosen (ER) bridge [20] and experimentally demonstrated in quantum entanglement of a single photon. [26] The collapse of the entire wave packet in one place then prevents its appearance in any other place.

Which particular part of the probability distribution function appears in the measurement depends by necessity upon the consequence of complex interfaces, aided by quantum entanglement, with the macroscopic detector possessing an enormous number of particles that are subject to irreversible thermal processes. Volumes have been written about the quantum measurement problem with a plethora of models for its solution. A consensus seems to be emerging on the efficacy of the environmental decoherence spearheaded by Zurek [27].

*Thus, the very weave of our universe appears to support the objective reality of the wave function, which represents a natural phenomenon and not just a mathematical construct. Furthermore, the nature of reality of the wave function described in this paper indeed requires the observed collapse of the wave function as well as a probabilistic outcome of measurement. This is offered as a proof of the ontology of the wave function presented in this paper.* Other confounding properties of a quantum particle also follow from the nature of the wave function described here.

As elaborated before, [10] the renowned uncertainty principle is in fact an inherent property of a wave packet. Due to dispersion, the wave packet would spread out rather quickly in position space. Since the wave packet is spread out before detection, the particle has the probability of being observed at more than one place. Thus the particle would appear to be present simultaneously at different places at the same time.

Also, because the particle is actually a (holistic) wave packet of the characteristics presented here, only the probability of the detection of its particle nature in a measurement can be predicted instead of a certainty as in classical physics. This is consistent with the customary assumption that the wave function is a function of probability amplitudes.

## 10. Conclusions

By the arguments presented in this paper, it should be reasonably evident that the wave--or more particularly the wave packet--associated with a material particle in the atomic dimensions is not just a fictitious mathematical construct for predicting results by solving the algorithm of quantum mechanics. It represents an objective reality, although not quite a classical one because of the inherently wave-like nature of the particle.

The principal aspect to bear in mind is that *only the sum total* of all the irregular disturbances in the quantum fields that comprise the travelling wave packet at any instant adds up to the mass, energy-momentum, charge and other conserved quantities of the particle. Consequently, it has to be taken all at the same time or not at all.

By this measure, the enigma of wave-particle duality is deciphered. Likewise, other apparently bizarre quantum behaviors such as the simultaneous existence of a quantum particle in more than one place, the uncertainty principle, achieving only the prediction of the probability rather than certainty of finding a particle in an experiment, the Born's rule, and wave function collapse can be given a satisfactory explanation thereby mitigating the perception of quantum weirdness that is so confounding to scientists and even more so to the general public. Now, nearly a century after the formulation of quantum mechanics, it is incumbent upon science to dispel the perception that the quantum core of our daily reality is of questionable realism.

**Acknowledgement**

The author wishes to thank Zvi Bern, Joseph Rudnick, and James Ralston for valuable discussions.